\documentclass[11pt,preprint]{aastex}
\shorttitle{ON THE TRANSIENT AXP XTE J1810--197}
\shortauthors{ERTAN $\&$ ERKUT}

\begin{document}

\title{ON THE X-RAY LIGHT CURVE, PULSED-RADIO EMISSION, AND SPIN FREQUENCY EVOLUTION OF THE TRANSIENT ANOMALOUS X-RAY PULSAR XTE J1810--197 DURING ITS X-RAY OUTBURST}
\author{\"{U}. ERTAN\altaffilmark{1} and M. H. ERKUT\altaffilmark{1}}

%\altaffilmark{1},
\affil{\altaffilmark{1}Faculty of Engineering and Natural Sciences, Sabanc\i\ University, 34956, Orhanl\i, Tuzla, \.Istanbul, Turkey}

\email{unal@sabanciuniv.edu, hakane@sabanciuniv.edu}

\begin{abstract}
We show that: (i) the long-term X-ray outburst light curve of the transient AXP XTE J1810--197 can be accounted for by a fallback disk that is evolving towards quiescence through a disk instability after having been heated by a soft gamma-ray burst, (ii) the spin-frequency evolution of this source in the same period can also be explained by the disk torque acting on the magnetosphere of the neutron star, (iii) most significantly, recently  observed pulsed-radio emission from this source coincides with the epoch of minimum X-ray luminosity. This is natural in terms of a fallback disk model, as the accretion power  becomes so low that it is not sufficient to suppress the beamed radio emission from XTE J1810--197.

\end{abstract}

\def\la{\raise.5ex\hbox{$<$}\kern-.8em\lower 1mm\hbox{$\sim$}}
\def\ga{\raise.5ex\hbox{$>$}\kern-.8em\lower 1mm\hbox{$\sim$}}
\def\be{\begin{equation}}
\def\ee{\end{equation}}
\def\ba{\begin{eqnarray}}
\def\ea{\end{eqnarray}}
\def\be{\begin{equation}}
\def\ee{\end{equation}}
\def\ba{\begin{eqnarray}}
\def\ea{\end{eqnarray}}
\def\m{\mathrm}
\def\d{\partial}
\def\R{\right}
\def\L{\left}
\def\a{\alpha}
\def\Mdot*{\dot{M}_*}
\def\Mdotin{\dot{M}_{\mathrm{in}}}
\def\Mdot{\dot{M}}
\def\Pdot{\dot{P}}
\def\Edot{\dot{E}}
\def\Omgdot{\dot{\Omega}}
\def\Omgastdot{\dot{\Omega}_{\ast }}
\def\Lin{L_{\mathrm{in}}}
\def\Rin{R_{\mathrm{in}}}
\def\rin{r_{\mathrm{in}}}
\def\rout{r_{\mathrm{out}}}
\def\Rout{R_{\mathrm{out}}}
\def\Omgast{\Omega _{\ast }}
\def\rA{r_{\mathrm{A}}}
\def\Ldisk{L_{\mathrm{disk}}}
\def\Lx{L_{\mathrm{x}}}
\def\dEb{\delta E_{\mathrm{burst}}}
\def\dEx{\delta E_{\mathrm{x}}}
\def\Bb{\beta_{\mathrm{b}}}
\def\Be{\beta_{\mathrm{e}}}
\def\Rc{\R_{\mathrm{c}}}
\def\dMin{\delta M_{\mathrm{in}}}
\def\dM*{\delta M_*}
\def\Teff{T_{\mathrm{eff}}}
\def\Tirr{T_{\mathrm{irr}}}
\def\Firr{F_{\mathrm{irr}}}
\def\Tcrit{T_{\mathrm{crit}}}
\def\Av{A_{\mathrm{V}}}
\def\ah{\alpha_{\mathrm{hot}}}
\def\ac{\alpha_{\mathrm{cold}}}
\def\p{\propto}
\def\m{\mathrm}

\section{INTRODUCTION}

Anomalous X-ray pulsars (AXPs) and soft gamma-ray repeaters (SGRs) are systems of young neutron stars identified mainly by X-ray luminosities ($10^{34}-10^{36}$ erg s$^{-1}$) much higher than their rotational powers indicated by their observed spin period and period derivatives (Woods \& Thompson 2005, Kaspi 2006). Brief soft gamma-ray bursts (\la 1 s) with super-Eddington luminosities observed from SGRs and some  AXPs led to the identification of these sources as magnetars (Duncan \& Thompson 1992). While the burst energetics are very likely to have magnetic origin, bursts themselves, being local events, do not require that the extended dipole component of the magnetic field has magnetar strength. Strong magnetic fields possibly in higher multipoles which decrease rapidly beyond the neutron star surface  is likely to be the source of these bursts. 

Fallback disk models (Chatterjee et al. 2000; Alpar 2001) present a self-consistent explanation for the period clustering, soft X-ray, optical and infrared (IR) light curves of AXPs and SGRs in their persistent and enhanced phases (Ek\c si \& Alpar 2003; Ertan \& Alpar 2003; Ertan \& Cheng 2004; Ertan et al. 2006; Ertan \& \c{C}al{\i}\c{s}kan 2006, Ertan et al. 2007). These explanations entail active, viscously dissipating disks around these systems. Recent observations of AXP 4U 0142+61 in the mid-IR bands with {\it Spitzer} Observatory (Wang et al. 2006) provided the first instance of a disk around an AXP.  Wang et al. (2006) attributed the optical and mid-IR data of this source to magnetospheric emission as proposed by magnetar models, and showed that the mid-IR data can be fitted with an irradiated passive disk. Detailed analysis of the overall data set including earlier detections in optical and near-IR bands strongly indicate the presence of an active disk around this source (Ertan et al. 2007). The difference between a passive and an active disk is that there is no viscous dissipation and therefore no mass inflow along a passive disk, while in an active disk, viscous dissipation works and leads to mass flow towards the inner disk and to the emission of X-rays through accretion onto the neutron star.  Both models expect a similar luminosity in the mid-IR bands which is emitted from the outer disk regions at which irradiation dominates dissipation. Nevertheless, estimates of these models differ in near IR and especially optical luminosities which, in the active disk model, are expected from the inner disk regions where the intrinsic dissipation becomes important. In a passive-disk model, the inner disk must be cut at a sufficiently large radius having a critical temperature below which temperatures are assumed to be not sufficient to sustain magnetorotational instability (MRI)(Balbus $\&$ Hawley 1991), as proposed by Wang et al.(2006) for the disk around AXP 4U 0142+61.  On the other hand, the minimum temperature required for MRI which leads to viscous dissipation is uncertain. Recent work by Inutsuka $\&$ Sano (2005) shows that MRI can be sustained in most regions of a protoplanetary disk even at temperatures as low as 300 K. Recently, Ertan et al. 2007 showed, by means of numerical fits to data, that the overall optical and IR dataset together with the X-ray luminosity of AXP 4U 0142+61 can be fitted with a single active, irradiated-disk model.  The upper limit for the dipole field strength estimated from these model fits is about $10^{13}$ G on the neutron star's surface (see Ertan et al. 2007 for details).        

The X-ray enhancement of SGR 1900+14 following its giant flare can be explained by the evolution of a disk after it has been pushed back and piled up at larger radii by the burst (Ertan \& Alpar 2003). This pushed-back disk model can also reproduce the long lasting, contemporaneous X-ray and IR enhancement light curve of AXP 1E 2259+586 for which the initial burst assumed by the model is likely to have remained below the detection limits (Ertan et al. 2006). In these models, reasonable fits to data are obtained by employing a single $\a$ viscosity (Shakura \& Sunyaev 1973) without invoking a disk instability. It might be  concluded that in the observed X-ray luminosity, $\Lx$, range of these and other  persistent AXPs and SGRs ($10^{34} - 10^{36}$ erg s$^{-1})$, fallback disks do not experience global disk instabilities.

Are these persistent sources already in their quiescent states, or do they have a lower viscosity state which will finally take them into quiescence with a sharp decrease in $\Lx$ after the mass-flow rate, $\Mdot$, at the inner disk has decreased to below some critical level? In this paper, we suggest that the known  persistent AXPs and SGRs are not in their quiescent states, and that the transient AXP XTE J1810$-$197 is a good example for an AXP that is now, following an outburst, making a transition  back into quiescence by means of a viscous disk instability. The observations of pulsed radio emission from this source fall in the epoch following the last X-ray observation by Gotthelf and Halpern (2007) (Fig 1). Our model X-ray light curve predicts that in this epoch the $\Lx$ remains even below the pre-outburst quiescent level. We propose that the pulsed-radio emission is a natural outcome of accretion rate decreasing below a critical value such that it is no longer sufficient to hinder pulsed-radio emission. In this picture,  all the other persistent AXPs and SGRs which are expected to be in the hot viscosity state at present would finally evolve into quiescence as their luminosity and thus the irradiation strength decrease below a critical level. The persistent and quiescent states would then correspond to two different viscosity states, and the AXPs and  SGRs could make transitions between these cold and hot states if the system undergoes a global instability triggered, for instance, by an energetic soft gamma-ray burst. Furthermore, persistent AXPs and SGRs are not expected  to show pulsed radio emission, because their accretion rates are much higher than the critical values below which pulsed-radio emission is allowed.             

Three of the AXPs (of which one is a candidate AXP) and one SGR  were observed to show transient behavior in their X-ray luminosity  (Kaspi 2006, Mereghetti et al. 2006, Israel  et al. 2007, Gotthelf and Vasisht 1998, Torii et al. 1998). Of these transient sources, XTE J1810$-$197 is a confirmed AXP with measured spin period and period derivative (Ibrahim et al. 2004). This source was discovered during the early decay phase of its X-ray outburst (Ibrahim et al. 2004). The $\Lx$  of XTE J1810$-$197 has still been decaying three years after its first detection, and  now is about two orders of magnitude less than  the observed maximum $\Lx$. This source was not observed around the onset of the X-ray outburst. Earlier archival data show that the $L_x$ prior to the outburst is also about two orders of magnitude fainter than the observed maximum luminosity (Gotthelf \& Halpern 2007).
During the first year of the decay phase of XTE J1810$-$197, $\Lx$ remained in the range characteristic of persistent AXPs. In this phase, the inner disk seems to be in a hot viscosity state, with temperatures above the critical level, likely to be the same as that operating in the persistent AXPs. In the present work, we show that for this first year, the  X-ray light curve of XTE J1810--197 can be simulated by a pure viscous evolution of a disk pushed back by a burst, but subsequently as $\Lx$ decreases below a critical level around $10^{34}$ erg s$^{-1}$ the light curve deviates from the characteristic viscous relaxation model curve.  This transition can be explained by a viscous disk instability leading the system into quiescence. 

The disk instability model for hydrogen disks is summarized and the possibilities for the instabilities of the fallback disks are discussed  in \S\ 2. In \S\ 3, we show that the overall decay curve of the transient AXP XTE J1810--197 can be reproduced by an irradiated active disk that undergoes a transition into quiescence below some critical X-ray luminosity that is predictable from the model fits. In \S\ 4 and \S\ 5, we discuss the X-ray light curve and its connection to pulsed radio emission respectively. In \S\ 6, we investigate the spin-frequency evolution of XTE J1810--197 during the post-burst fading of the X-ray flux. Our results are summarized in \S\ 7.

\section{INSTABILITY IN A FALLBACK DISK }

The viscous disk instability model (DIM)  for hydrogen disks (see e.g. Lasota 2001 for a review) works briefly as follows: At a given radial distance, the disk is in one of two states: either a hot or a cold stable viscosity state. For the range of effective temperatures, $\Teff$, corresponding to partial ionization of hydrogen ($\sim 6000-10000$ K) the disk is locally unstable. Above this temperature range, in the hot state, the mass-flow rate, $\Mdot$, is high due to high viscosities in this state. If this local $\Mdot$ is higher than that supplied from the outer radii, surface density, $\Sigma$,  and $\Teff$ decrease. When $\Teff$ drops below the critical temperature, $\Tcrit$,
the system enters the cold state with the same $\Sigma$, but with much lower $\Mdot$ and  $\Teff$ than those of the hot state.  
In the cold phase, because of low viscosities, the matter accumulates in this radius causing $\Sigma$  and $\Teff$ to increase gradually. Finally, $\Teff$ exceeds $\Tcrit$ and the system locally enters the hot regime, thereby completing the limit cycle behavior. This local transition  to the hot state causes a heating front to propagate towards larger and smaller radii increasing the temperatures. If this propagation of the heating front 
takes a large portion of the disk to the hot state, enhanced $\Mdot$ towards the inner disk and thus onto the neutron star is observed as an X-ray outburst in the light curve. Subsequently, when the effective temperature drops below the critical temperature a cooling front from the outer disk propagates radially inward, characterizing the light curve as the disk evolves into quiescence.  

When X-ray irradiation is effective the cooling front cannot propagate inward rapidly with local sound speeds as it does in the absence of irradiation. Instead, it propagates inwards slowly, governed by the X-ray irradiation flux from the neutron star.

Although the composition of fallback disks is uncertain, it is likely that they all have similar, probably metal-rich, compositions and accordingly similar critical temperatures  for viscous disk instabilities to be triggered. 
It is possible that a fallback disk could have more than one critical temperature depending on the composition and the ionization temperatures of the constituent elements. In this case, the minimum critical temperature that could create a disk  instability is expected to be the temperature that corresponds to the minimum ground-state ionization energy among  all components of the chemical composition. In \S\ 3, from the model fits, we try to constrain the possible range of critical temperatures responsible for the transition of XTE J1810$-$197 into the quiescent state.

\section{NUMERICAL MODEL}

The transient AXP XTE J1810$-$197 was not observed around the onset of the
X-ray outburst. We assume that the outburst was triggered by a soft gamma-ray burst by pushing the inner disk to larger radii. For the numerical calculations of the diffusion equation, we represent the initially  pushed-back disk by a Gaussian surface density distribution at the inner disk, $\Sigma (r,t=0)=\Sigma _{\max}\exp \lbrack -(r-r_{0})^{2}/(\Delta r)^{2}] $ (inner pile-up), interpolated to a power-law distribution representing the outer extended disk, $\Sigma = \Sigma_0 (r_0/r)^p$, where $r_0$ and $\rin$ are the center of the Gaussian and the inner disk radius respectively. For all the model curves presented in Figure 1, $\rin=1.0\times 10^9$ cm,  $\rout=5.0\times 10^{12}$ cm, $r_0=6.0\times 10^9$ cm, $\Sigma_{\m{max}}= 120$ g cm$^2$,  $\Sigma_0= 8.0$ g cm$^2$, $\Delta r= 6.0\times 10^8$ cm, and $p= 0.75$. Model curves do not depend on the details of the Gaussian (or other) model distributions (see Ertan et al. 2006 for details).       
We follow the same method as that applied to 1E 2259+586 and SGR 1900+14. 
The difference here is that we now introduce a critical effective temperature $\Tcrit$ above which the disk is locally in the hot state with
the $\a$ parameter $\ah =0.1$. At the radial grid points, with  $\Teff < \Tcrit$  we set $\a= \ac$  and  obtain reasonable fits to data  with $\ac \simeq 0.023$. The values of $\ah$ and $\ac$ are similar to those employed in disk models of soft X-ray transients and dwarf novae. In our calculations, we employ electron scattering opacities, since the exact metal composition and the corresponding opacities are not available. It should be kept in mind that  the $\a$ parameters of the model include uncertainties in opacities. In our model, we take the mass accretion from the fallback disk onto the neutron star as the source of the X-ray luminosity. Our aim is to see whether the model can reproduce the long-term behavior of the X-ray light curve. In this model, X-ray emission is determined by equating the mass-inflow rate at the innermost disk to the accretion rate onto the neutron star. The resultant energy spectra and pulsed fractions depend completely on the details of emission mechanisms close to  neutron star (accretion geometry, Comptonization effects, outflows etc.) which are out of the scope of the present work. 

Our best model fits with and without disk instability for XTE J1810$-$197 are presented in Figure 1. The disk model with instability does not by itself specify the value of the critical effective temperature due to the existence of another parameter, the irradiation efficiency, $C$, which is related to the irradiation flux as $\Firr= C \Mdot_* c^2/4 \pi r^2$, where $c$ is the speed of light, $\Mdot_*$ is the accretion rate on to the neutron star, and $r$ is the radial distance in the disk.
The value of  $C$ estimated from the simultaneous X-ray and IR observations in the $K_s$ band (Rea et al. 2004) is 2.5 $\times 10^{-5}$, 
which is about 25 \% of that estimated for 1E 2259+586 and 4U 0142+61 in their persistent states (Ertan \& \c{C}al{\i}\c{s}kan 2006). The observed flux in the H band (Rea et al. 2004) is exactly equal to the flux in this band estimated  from a non-irradiated disk, indicating that the middle disk region emitting in the H band was probably shielded by the hot innermost disk regions as the cooling front was propagating towards the inner disk. $K_s$ band emission during the IR observations is expected from a radius of about $8 \times 10^{10}$ cm which is nearly two times the radius of the disk region emitting mainly in the H band. The disk region emitting in the $K_s$  band seems to be partly shielded also, considering low irradiation efficiency as compared to the other AXPs. Alternatively, the inclination angle of the disk in XTE J1810--197 could be much larger than those of the other persistent AXPs, leading to irradiation efficiencies larger than those estimated above. Taking these possibilities into account,  in order to constrain  the critical temperature of the disk, we try three different irradiation efficiency parameters for XTE J1810--197: the low efficiency $C=2.5 \times 10^{-5}$ estimated by comparing the X-ray flux to the flux in the $K_s$-band emission, $C=5.6 \times 10^{-4}$ that is around the maximum of the estimated irradiation efficiencies for all the transient AXPs (Ertan \& \c{C}al{\i}\c{s}kan 2006 ), and  $C=5.7 \times 10^{-5}$ that is between these extremes. We obtained, from the model fits, a critical effective temperature range of $\sim$ 1200$-$2500 K limited by the  smallest and largest $C$ values, respectively.       

\section{THE X-RAY LIGHT CURVE}

We would like to present a prediction of the disk model at the end of the decay curve of the XTE J1810$-$197 after the X-ray observations presented here. From our model curves, we see that the inner disk does not secularly settle down to the quiescent state with a continuous decay in the X-ray luminosity, $\Lx$. Instead, for a large range of the critical temperatures corresponding to the plausible range of the irradiation parameter, the light curve $\Lx(t)$ exhibits a dip structure, before it reaches the steady-like quiescent level (see Fig. 1). The reason for this is the surface density gradient created between the inner and outer regions of the cooling front as it propagates towards the inner disk. As the cooling front comes closer to the inner disk, the mass-depletion rate in the inner region is faster than in the outer region of the front. The resulting surface density gradient leads to an abrupt increase in $\Mdot$ at the innermost disk. This is observed as a dip-like structure in the X-ray light curve. This activity causing density fluctuations around the cooling front is actually an ongoing process during the earlier overall decay phase. However, initially these surface density gradients are far from the inner disk and smoothed out along the way to the inner disk radius, and therefore do not significantly modify the X-ray light curve. Once this effect has reached the inner disk, it already signals the end of the secular decay. Increasing $\Mdot$ from the cold to hot inner regions due to these density gradients increases $\Lx$ (turn-up from the dip) and the resulting increse in the irradiation strength takes some inner part of the currently cool region into the hot zone. But, now, the density gradients at the inner disk are smoothed out once again and $\Lx$ converges to its quiescent level.  For our set of model parameters,  dip structures and the following X-ray light curves expected  from our model are presented in Figure 1. It is remarkable that following the rise from the minimum, all the model curves with different critical temperatures converge to a similar mean $\Lx$. Note that during a time span of about a year around the minimum of the model curve, $\Lx$ and $\Mdot$ of this source are below their pre-outburst quiescent levels. It is very significant in that observations of pulsed radio emission for XTE J1810-197  (Camilo et al. 2007) also coincide with this period.

\section{PULSED RADIO EMISSION} 

Pulsed radio emission is expected from a polar gap for an isolated radio pulsar (see e.g. Lyne and Smith 1998) as a result of the streaming bunches of the electrons along the field lines. For an accreting neutron star, pulsed radio emission is quenched beyond a critical accretion power likely to be around the gap power $\Edot\simeq I \Omgast \Omgastdot$ producing the pulsed radio emission. $I\simeq 10^{45}$ g cm$^2$ is the moment of inertia of the neutron star, $\Omgast$ and  $\Omgastdot$ are the spin angular frequency and its time derivative respectively. Using the observed spin period $P = 2 \pi / \Omega =5.54$ s and period derivative $\Pdot\simeq 1\times 10^{-11}$ of XTE J1810-197 (Ibrahim et al. 2004), we obtain $\Edot\simeq 2\times 10^{33}$ ergs s$^{-1}$. The X-ray luminosity during the epoch of the observations of the pulsed-radio emission estimated from the observed X-ray flux of this source is 
$\Lx ~\la~ 1 \times 10^{33} d_{3.3}$ ergs s$^{-1}$ where $d_{3.3}$ is the distance to the source in units of 3.3 kpc (Camilo et al. 2006).

XTE J1810--197 is the only AXP which was observed to emit pulsed-radio emission (Camilo et al. 2006; 2007). In the epoch of the pulsed-radio observations ($53880 - 54127$ MJD) that followed the most recent X-ray observations with XMM (Gotthelf $\&$ Halpern 2007), X-ray data is not available. This is the epoch in which our model predicts that the X-ray luminosity of the source remains below its  quiescent level prior to the outburst. It is remarkable that clear radio-pulse profiles become noisy as the model X-ray luminosity increases from its minimum after the end of the decay phase (see Fig. 1). We propose that during the time interval in which the source shows pulsed-radio emission, the accretion rate is below a critical level such that accretion power is insufficient to close the magnetospheric gap completely and the source could produce beamed radio emission. The accretion and gap powers estimated above are consistent with our prediction. As the accretion rate and thereby the X-ray luminosity increase, returning to the pre-outburst level, the radio-pulse profiles become noisy and gradually disappear. Persistent AXPs and SGRs are not expected to show pulsed-radio emission simply because their accretion power is sufficient for suppressing the beamed radio emission.

\section{SPIN EVOLUTION OF THE TRANSIENT AXP XTE J1810--197\label{sfe}}

The torque acting on a magnetized neutron star interacting with an accretion
disk can be written in general as 
\be
N_{\ast }=j\dot{M}(GM_{\ast }r_{\mathrm{in}})^{1/2}, \label{1}
\ee
where $\dot{M}$ 
is the mass-inflow rate in the disk, $M_{\ast }$
is the mass of the neutron star, and $j$ is the nondimensional torque as a
function of the fastness parameter $\omega _{\ast }\equiv \Omega _{\ast
}/\Omega _{\mathrm{K}}(r_{\mathrm{in}})$, $\Omega _{\ast }$ being the
angular spin frequency of the neutron star and $\Omega _{\mathrm{K}%
}(r)=(GM_{\ast }/r^{3})^{1/2}$ the Keplerian angular velocity (see Ghosh \&
Lamb 1979; Erkut \& Alpar 2004). The inner disk radius $r_{\mathrm{in}}$ can
be taken, within an order of magnitude, to be the Alfv\'{e}n radius, $r_{%
\mathrm{A}}=\dot{M}^{-2/7}(GM_{\ast })^{-1/7}\mu _{\ast }^{4/7}$, where $\mu
_{\ast }$ is the dipole magnetic moment of the neutron star.

When the star rotates faster than the magnetically coupled inner disk ($%
\omega _{\ast }>1$), that is, when the Alfv\'{e}n radius is larger than the
co-rotation radius $r_{\mathrm{co}}\equiv (GM_{\ast }/\Omega _{\ast
}^{2})^{1/3}$, the classical assumption is that the neutron star acts as a
propeller throwing most of the incoming disk matter out of the system
(Illarionov \& Sunyaev 1975). However, this picture cannot be entirely
correct because it is not energetically self-consistent; instead, disks that
truncate near the co-rotation radius may allow accretion with spin-down of
the star (see Rappaport et al. 2004).

In the spin-down with accretion regime ($\omega _{\ast }>1$), there should
be a transition region within which the disk matter is brought into
co-rotation with the neutron star as a result of angular momentum exchange
between the magnetosphere and the disk. Taking the effective star-disk
coupling radius to be $r_{\mathrm{A}}$, the magnetospheric torque applied by
the disk on the star can be calculated as%
\begin{equation}
N_{\mathrm{mag}}=-\int_{r_{\mathrm{co}}}^{r_{\mathrm{A}}}r^{2}B_{z}B_{\phi
}^{+}dr,  \label{gte}
\end{equation}%
where $B_{z}\simeq -\mu _{\ast }/r^{3}$ is the vertical component of the
dipolar magnetic field of the neutron star and $B_{\phi }^{+}$ is the
azimuthal component of the magnetic field on the surface of the disk (see
Erkut \& Alpar 2004). The magnetospheric torque in equation (\ref{gte})
represents the flux of angular momentum transferred to the magnetic field of
the neutron star as the disk matter couples with the field lines at $r=r_{%
\mathrm{A}}$ and accretes onto the star over the range $r_{\mathrm{co}}\leq
r\leq r_{\mathrm{A}}$ (see, e.g., Dai \& Li 2006). The azimuthal component
of the magnetic field can be written as $B_{\phi }^{+}=\gamma _{\phi }B_{z}$%
, where $\gamma _{\phi }$ is the azimuthal pitch of order unity and $\gamma
_{\phi }>0$ for $r\gtrsim r_{\mathrm{co}}$ (see Wang 1995). Assuming that
the star-disk interaction is of magnetospheric origin ($N_{\ast }=N_{\mathrm{%
mag}}$), it follows from equation (\ref{gte}) that the nondimensional torque is $j=(\gamma _{\phi }/3)(1-\omega _{\ast }^{2})$, where $\omega _{\ast }=(r_{\mathrm{A}}/r_{\mathrm{co}})^{3/2}$ is the
fastness parameter. The second term in the nondimensional torque $j$ is proportional to $\omega _{\ast }^{2} \propto \Mdot^{-6/7}$. Substituting $j$  into equation (\ref{1}) we see that $N_{\ast }\propto j \Mdot^{6/7}$ becomes almost independent of $\Mdot$ since the second term in $j$ dominates the spin-up term as the neutron star spins down while accreting. The second term is
reminiscent of the spin-down torque of the subsonic propeller phase
described by Davies et al. (1979). If the star is close to rotational
equilibrium with the disk, the nondimensional torque changes character and behaves as $j\propto (1-\omega _{\ast })$ for $\omega _{\ast }\simeq 1$ as
expected (see, e.g., Ertan et al. 2007).

Our model fits to the X-ray outburst light curve  of XTE J1810--197 are
consistent with the assumption  that all of the incoming disk matter
accretes onto the neutron star. Taking the X-ray
luminosity of the source to be an accretion luminosity, the time variation
of the mass inflow rate $\dot{M}$ in the inner disk can be incorporated with
the corresponding changes in $r_{\mathrm{A}}$ and $\omega _{\ast }$ to
estimate the spin-frequency evolution of the neutron star. The frequency
evolution of the neutron star is obtained using $N_{\ast }=2\pi I_{\ast }%
\dot{\nu}_{\ast }$, where $I_{\ast }\simeq 10^{45}$
g cm$^{2}$ and $\nu _{\ast }\simeq 0.18$ Hz are the neutron star's moment of
inertia and spin frequency, respectively. Figure 2 shows the best fit of our
torque model to the frequency data from Ibrahim et
al. (2004) and Gotthelf \& Halpern (2007) for $\gamma _{\phi }=3/2$. Our model curve for the spin-frequency evolution of XTE J1810--197 (Fig.2) corresponds to a surface dipole magnetic field of $B_{\ast }\simeq
2\times 10^{12}$ G ($\mu _{\ast }\simeq 10^{30}$ G cm$^{3}$) on the poles of
a neutron star of mass 1.4 $M_{\odot }$ and radius 10$^{6}$ cm. As $\dot{M}$  decreases, reflecting a decrease in the X-ray
flux, $\rA$ and the fastness parameter $\omega _{\ast }$
increase. In the long-term spin evolution of the neutron star, XTE
J1810--197, as a transient AXP, is farther away from the rotational
equilibrium with the disk as compared with its persistent cousins. In this phase, the observed spin-down torque remains almost constant, independent of $\dot{M}$ as estimated from our torque expression. Note that as the accretion rate drops further it is likely that the neutron star propels most of the inflowing disk matter out of the system. In this case, it is expected that the spin-down torque acting on the neutron star becomes less efficient than that observed in the present phase.   

\section{DISCUSSION AND CONCLUSIONS}

We have firstly shown that the X-ray outburst light curve of the transient AXP XTE J1810$-$197 lasting for about three years can be reproduced by the evolution of an irradiated disk that undergoes a transition into quiescence by means of a viscous disk instability as the X-ray luminosity decreases below a critical limit less than around $10^{34}$ erg s$^{-1}$ which is near the minimum of the observed luminosity range of AXPs and SGRs. We have assumed that the inner disk is initially pushed back by a missed soft gamma-ray burst that
took a large portion of the inner disk into a hot viscosity state which is likely the state prevailing in the disks of persistent AXPs and SGRs.
The mass-flow rate, $\Mdot$,  provided by the extended disk of XTE J1810$-$197 with surface densities likely to be lower than those of persistent sources is not sufficient for the inner disk to be kept in the hot state for longer periods by fuelling the X-ray irradiation through accretion. After the X-ray luminosity has decreased below a critical level, the system evolves into quiescence, by means of a viscous disk instability. Observations indicate that the source was also in the quiescent state prior to the X-ray outburst. Unlike XTE J1810$-$197, persistent sources AXP 1E 2259+586 and SGR 1900+14 were already in the hot state with 
$\Lx ~\ga ~10^{34}$ erg s$^{-1}$ before their enhancements and their X-ray luminosities remained above the critical level during the outburst decay as well. The X-ray luminosity  of these and other persistent AXPs and SGRs are also expected to decrease slowly as the disk surface densities decrease gradually through expansion and accretion in the persistent phase. They will also finally evolve towards the quiescent regime by means of a disk instability as their accretion rate falls below critical values. Quiescent sources, on the other hand, could make an upward transition to the hot state when a viscous instability is triggered by an energetic gamma-ray burst. 
               
Secondly, we have shown that the magnetospheric torque applied by the disk on the star is in agreement with the spin-frequency evolution of this transient AXP (see Fig. 2). Along the X-ray outburst decay, XTE J1810$-$197 is in the accretion regime with spin-down. In this phase, the system is slowly receding from the rotational equilibrium between the disk and the magnetosphere, while most of the mass flowing towards the disk is being accreted onto the neutron star. Our torque model
is almost independent of the mass-inflow rate $\Mdot$ to the inner disk in this period.  We have obtained the model curve for spin evolution directly from the evolution of the mass-inflow rate corresponding to the model X-ray light curve presented in Fig. 1 and by adjusting the field strength accordingly. The model curve given in Fig.2 is obtained with a dipolar magnetic field of $B_{\ast }\simeq 2\times 10^{12}$ G. Below some critical value of $\Mdot$, most of the inflowing matter is propelled out of the system, and the functional form of the torque is expected to change. In this regime the torque would probably be less efficient than the model torque we employ here. The dependence of the torque on $\Mdot$ also changes as the system approaches rotational equilibrium (see \S\ 6), and the spin-up torque due to accretion is no longer negligible in comparison with the spin-down torque.

Finally, we have proposed that the pulsed-radio emission from XTE J1810--197 could be accounted for by the decreasing accretion rate below a critical level which switches on the beamed radio emission from the polar caps. Our model predicts a dip-like structure at the end of the decay phase such that the X-ray luminosity remains even below the pre-outburst quiescent level for about one year. This period matches the epoch of observations of pulsed-radio emission from this transient AXP (Camilo et al. 2007). This interpretation of the pulsed radio emission together with the model predictions for the X-ray light curve could be tested by future radio and X-ray  observations of XTE J1810--197 and other transient AXPs and SGRs. Transient AXPs can show up as radio pulsars when their disks make a transition to the cold state and the X-ray luminosity drops sufficiently such that their accretion power can no longer suppress the beamed radio emission.   

A detailed analysis on the torque evolution and the beamed radio emission from AXPs and SGRs will be the subject of our future work.

\acknowledgements

We acknowledge research supports from T\"{U}B\.{I}TAK (The Scientific and Technical Research Council of Turkey) through the grant 107T013 and from the Sabanc\i\ University Astrophysics and Space Forum. We thank Ali Alpar, Feryal \"{O}zel, Dimitrios Psaltis, Ersin G\"{o}\u{g}\"{u}\c{s}, Emrah Kalemci and Yavuz Ek\c{s}i for useful discussions.  We thank E. Gotthelf and J. Halpern for providing X-ray data of XTE J1810--197. This work has been supported by the  Marie Curie FP6 Transfer of Knowledge Project ASTRONS, MKTD-CT-2006-042722.  

%\clearpage

\clearpage
\begin{figure}
\vspace{-10 cm}
\hspace{-4 cm}
\includegraphics{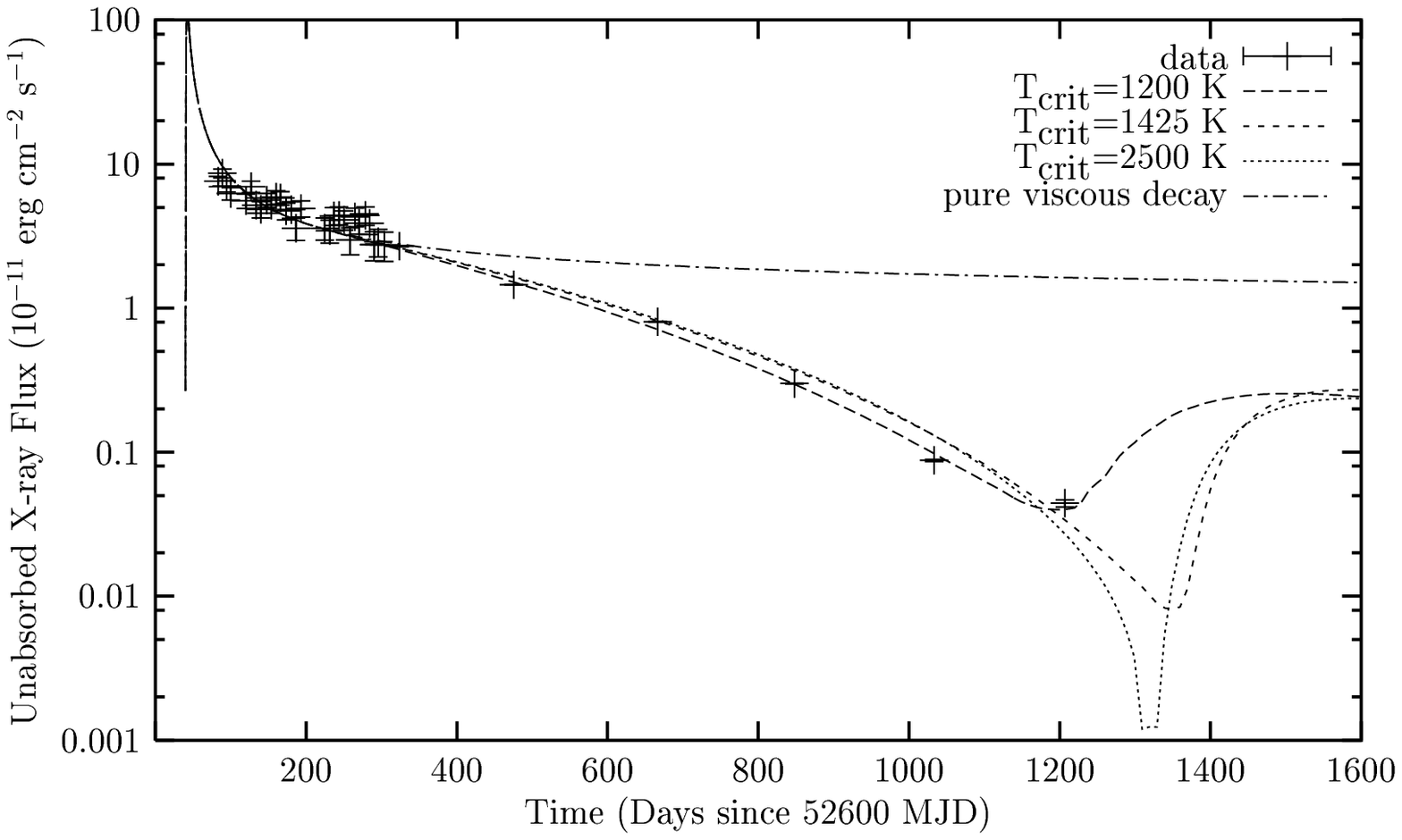}
\vspace{-11 cm}
\caption{Values of the irradiation parameter $C$ is $2.5 \times 10^{-5}, 5.7 \times 10^{-5}$, and $5.6 \times 10^{-4}$ for $\Tcrit=$ 1200, 1425, and 2500 K    respectively. Initial surface density distributions,  are the same for all model curves including the pure viscous evolution (see \S\ 3 for model parameters). The data points were taken from Ibrahim et al. (2004) and  Gothelf and Halpern (2007)}
\end{figure}

\clearpage
\begin{figure}
\vspace{-10 cm}
\hspace{-4 cm}
\includegraphics{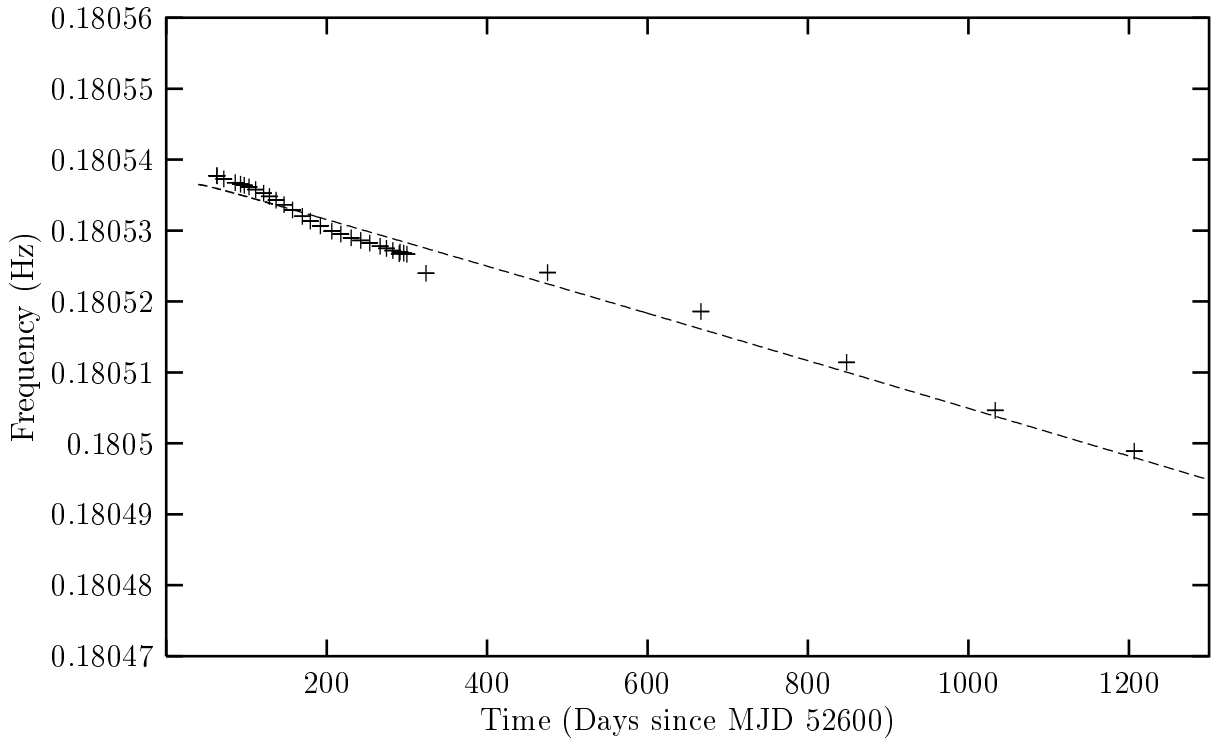}
\vspace{-11 cm}
\caption{Torque-model fit for the spin-frequency evolution of the XTE J1810--197, see \S\ 6  for details. The data points for the first 250 days were obtained from the polynomial fit to frequency data by Ibrahim et al.  (2004) and the last seven data points were taken from Gothelf and Halpern (2007). The data points do not show the error bars.}
\end{figure}

\end{document}